\documentstyle[aps,12pt,preprint]{revtex}

\renewcommand{\vec}[1]{\mbox{\boldmath $#1$}}

\begin{document}
\preprint{}
\title{How harmonic is dipole resonance of metal clusters?}
\author{K. Hagino}
\address{Institute for Nuclear Theory, Department of Physics, 
University of Washington, \\ Seattle, WA 98195, USA \\ and \\
Department of Physics, Tohoku University, Sendai 980--8578, Japan}

\maketitle

\bigskip

\begin{abstract}

We discuss the degree of anharmonicity of dipole plasmon 
resonances in metal clusters. We employ the time-dependent 
variational principle and show that the relative shift 
of the second phonon scales as $N^{-4/3}$ in energy, 
$N$ being the number of particles.
This scaling property coincides with that for nuclear giant 
resonances. 
Contrary to the previous study based on the boson-expansion 
method, the deviation from the harmonic limit is found to be 
almost negligible for Na clusters, the result being consistent 
with the recent experimental observation. 

\end{abstract}



Collective phonon excitations are common phenomena 
in fermionic many-body systems. 
In alkali metal clusters, a very strong dipole 
plasmon resonances have been observed in the electromagnetic 
response, which are interpreted as collective vibrational 
excitations of an electron gas against a background ions 
\cite{H93}. 
These resonances are well described in the 
random phase approximation (RPA) as a particle-hole 
excitaion, which assumes the harmonic nature of the 
vibrations \cite{BB94}. 
It may thus be a natural idea to expect the existence of 
multiple plasmon excitations. 
It is worth mentioning that double phonon excitations have 
been observed in a similar phenomenon in nuclear 
physics, i.e. giant dipole resonance \cite{ABE98,E94}. 

One interesting question is whether the dipole plasmon 
resonance in metal clusters is harmonic enough to allow 
multiple excitations. 
Catara {\it et al.} used the boson-expansion method 
to discuss two-plasmon excitations in metal clusters \cite{CCG93}. 
They claimed that anharmonic effects are quite large. 
On the other hand, a comparison between a jellium-RPA calculation and 
the result of time-dependent local density approximation (TDLDA) \cite{YB96} 
suggests that the anharmonic effects are very small for Na clusters. 
Recently an experiment was performed which addresses the anharmonic 
properties of plasmon resonances of alkali metal clusters \cite{SKIH98}. 
In this experiment, a large intensity 
of doubly and triply charged fragments were obserbed in the 
charge distribution of photofragment of Na$_{93}^+$. 
The ionization energy of this cluster lies between the energy 
of single and double plasmon resonances, and the photon energy 
was set to be slightly larger than the one plasmon energy. 
The ionization was thus energetically possible only if at least 
two photon are absorbed. The observed ionization was 
interpreted as electron emission via multiple plasmon states, suggesting 
a very small anharmonicity, which does not support the prediction 
by Catara {\it et al.} This picture was later confirmed 
theoretically, although a significant correction from a direct 
two photon absorbtion was reported \cite{BGM99}. 

The aim of this paper is to show that the anharmonic effects are indeed 
very small for the dipole plasmon resonance in metal clusters, 
contrary to the prediction by Catara {\it et al.}
To this end, we apply the variational principle for the time-dependent 
Schr\"odinger equation. 
The time-dependent variational principle was recently applied to 
large amplitude collective motions to discuss 
anharmonic properties of nuclear giant resonances \cite{BF97}. 
Its applicability has been tested on a solvable model in Ref. \cite{BBH99}. 
The time-dependent variational approach allows one to estimate 
relatively easily the energy shift of double phonon state with an 
analytic formula. 
As we will see, it has the same scaling 
law concerning the number of particle for both nuclear and cluster cases, 
although the range of the interaction is very different for these 
two systems. 

Consider a system where $N$ electrons interact with each other 
in a positively charged ionic background whose density is given by 
$\rho_I(\vec{r})$. The Hamiltonian for this system is given by  
\begin{equation}
H=\sum_{i=1}^N\frac{\vec{p}_i^2}{2m}+\frac{1}{2}\sum_{i\neq j}
\frac{e^2}{\left|\vec{r}_i-\vec{r}_j\right|} + 
\sum_{i=1}^NV_I(\vec{r}_i),
\label{H}
\end{equation}
where $m$ is the electron mass. 
$V_I(\vec{r})$ describes the interaction between the electrons 
and the ionic background. It is given by 
\begin{equation}
V_I(\vec{r})=-e^2\int d\vec{r}_I
\frac{\rho_I(\vec{r}_I)}{\left|\vec{r}-\vec{r}_I\right|},
\end{equation}
and satisfies the Poisson equation
\begin{equation}
\nabla^2V_I(\vec{r})=4\pi e^2\rho_I(\vec{r}).
\label{poisson}
\end{equation}
As in Ref. \cite{BF97}, 
we assume the following wave function to discuss the non-linear dynamics 
of the dipole plasmon resonance of this system. 
\begin{equation}
|\Psi_{\alpha\beta}> = e^{i\alpha(t)Q}|\Psi_{\beta}> = 
e^{i\alpha(t)Q}e^{m\beta(t)[H,Q]}|\Psi_0>.
\end{equation}
Here $|\Psi_0>$ is the ground state wave function. 
The operator $Q$ is the dipole field given by
\begin{equation}
Q=\sum_{i=1}^Nz_i.
\label{dipole}
\end{equation}
The time evolution of the variables $\alpha(t)$ and $\beta(t)$ is 
determined according to the time-dependent variational principle, 
\begin{equation}
\delta\int dt<\Psi_{\alpha\beta}|i\partial_t-H|\Psi_{\alpha\beta}>=0.
\end{equation}
This leads to the following two coupled equations \cite{BF97}
\begin{eqnarray}
\dot{\beta}&=&\frac{\alpha}{m}, \\
\dot{\alpha}\partial_{\beta}<\Psi_{\beta}|Q|\Psi_{\beta}> 
&+&\partial_{\beta}<\Psi_{\beta}|H|\Psi_{\beta}> +
\frac{\alpha^2}{2}\partial_{\beta}<\Psi_{\beta}|[Q,[H,Q]]|\Psi_{\beta}> =0.
\end{eqnarray}

In order to find the phonon frequency, we requantise these equations of 
motion with a Hamiltonian formulation. One convenient choice of the canonical 
transformation is \cite{BF97}
\begin{eqnarray}
\beta&\to&q=\beta, \\
\alpha&\to&p=m\alpha<\Psi_{\beta}|[Q,[H,Q]]|\Psi_{\beta}>,
\end{eqnarray}
together with the Hamiltonian of 
\begin{equation}
{\cal H}=\frac{p^2}{2M(\beta)} + U(\beta),
\end{equation}
where the inertia and the potential are given by 
\begin{eqnarray}
M(\beta)&=&m^2<\Psi_{\beta}|[Q,[H,Q]]|\Psi_{\beta}>,  \\
U(\beta)&=& <\Psi_{\beta}|H|\Psi_{\beta}>,
\end{eqnarray}
respectively. 
For the dipole field (\ref{dipole}), the inertia is easily 
evaluated to be $M(\beta)=mN$. 
After dropping the constant term, the collective potential is calculated as 
\begin{equation}
U(\beta)=\int dxdydz V_I(x,y,z-\beta)\rho_0(x,y,z)
=\int dxdydz V_I(x,y,z)\rho_0(x,y,z+\beta),
\label{pot}
\end{equation}
$\rho_0$ being the ground state density. 
To derive Eq. (\ref{pot}), we have used the transformation
\begin{equation}
e^{-m\beta[H,Q]}(x_i,y_i,z_i)e^{m\beta[H,Q]}
=(x_i,y_i,z_i-\beta)
\end{equation}
for the dipole field, and 
the fact that both the kinetic energy 
and the Coulomb interaction among the electrons 
in the Hamiltonian (\ref{H}) are translational 
invariant \cite{B89}. 
Since we are interested in the leading order correction to the 
harmonic limit, we expand the ground state density in Eq. (\ref{pot}) 
in terms of $\beta$. Accordingly, we express the collective potential as 
\begin{equation}
U(\beta)=U_0 + \frac{k}{2}\beta^2+\frac{k_4}{4}\beta^4 + \cdots.
\end{equation}
The linear term in the expansion vanishes because of the 
stability of the ground state, and the third order term drops if 
the spherical symmetry is assumed for the ground state density. 
Using the Poisson equation (\ref{poisson}), the coefficients $k$ and 
$k_4$ are evaluated as
\begin{eqnarray}
k&=&\frac{16\pi^2}{3}e^2\int^{\infty}_0r^2dr\rho_I(r)\rho_0(r), 
\label{k} \\
k_4&=&-\frac{16\pi^2}{54}e^2
\int^{\infty}_0r^2dr\frac{d\rho_I(r)}{dr}\frac{d\rho_0(r)}{dr},
\label{k4}
\end{eqnarray}
respectively. Here we have assumed that the ionic density $\rho_I$ 
has the spherical symmetry. The expression for $k$, Eq. (\ref{k}), was 
first derived by Brack in the context of the sum rule approach 
\cite{B89,B93}. 

Equations (\ref{k}) and (\ref{k4}) are general expressions and valid for 
any form of the ionic and the electronic density distributions, as long as 
they are spherical. 
In order to get a more transparent formula, we further simplify them 
by using the jellium approximation. 
Here the ionic charge density is uniform in a sphere of radius 
$R=r_sN^{1/3}$, where $r_s$ is the Wigner-Seitz radius. 
We thus assume 
\begin{equation}
\rho_I(r)=\frac{3}{4\pi r_s^3}\theta(R-r),
\label{jellium}
\end{equation}
where $\theta$ is the theta function. 
Substituting Eq. (\ref{jellium}) to Eqs. (\ref{k}) and (\ref{k4}), we obtain
\begin{eqnarray}
k&=& \frac{Ne^2}{r_s^3}, \\
k_4&=&\frac{4\pi}{18}\frac{e^2}{r_s}N^{2/3}
\left.\frac{d\rho_0(r)}{dr}\right|_{r=R}. 
\label{k4-2}
\end{eqnarray}
Here we have 
neglected the effect of spillout of the electron density 
outside the ionic background, which is order of $1/N$. 
To get an analytic expression for $k_4$, we approximate the electronic 
density by the error function as \cite{BE85}
\begin{equation}
\rho_0(r)=\frac{3}{4\pi r_s^3}\mbox{erfc}\left(\frac{R-r}{a}\right),
\label{error}
\end{equation}
where $a$ is the surface diffuseness parameter for the electronic 
density. 
Substituting this density to Eq. (\ref{k4-2}), we finally obtain 
\begin{equation}
k_4=-\frac{e^2N^{2/3}}{3a\sqrt{\pi}r_s^4}.
\end{equation}

We now requantise the collective Hamiltonian ${\cal H}$ and 
obtain the phonon frequency. 
The frequency in the harmonic limit is given by
\begin{equation}
\omega_0=\sqrt{\frac{k}{M}}=\sqrt{\frac{e^2}{mr_s^3}}, 
\label{Mie}
\end{equation}
which coincides with the Mie frequency. 
The leading correction to the harmonic limit is given by \cite{BF97,BBH99}
\begin{equation}
E_n = n\omega_0 -\frac{3k_4}{8k^2}n^2\omega_0^2
= n\omega_0 -\frac{1}{8\sqrt{\pi}}\frac{r_s^2}{ae^2}N^{-4/3}n^2\omega_0^2.
\end{equation}
Taking the second derivative, the energy shift of the double 
phonon state is found to be
\begin{equation}
\Delta^{(2)}E=E_2-2E_1-E_0=
-\frac{\omega_0^2}{4\sqrt{\pi}}\frac{r_s^2}{ae^2}N^{-4/3}.
\label{shift}
\end{equation}
Note that the dependence of $\Delta^{(2)}E/\omega_0$ on 
$N$ is the same as that for the nuclear giant resonances \cite{BF97}. 
Expressing it as a dependence of the volume of the 
system $N\sim L^3$, it is also the same as that for 
the photon spectrum in a small cavity \cite{BBH99}. 
All of them scale as $N^{-4/3}\sim L^{-4}$. 

Let us now estimate numerically the shift of the frequency for 
Na clusters. Using the 
Wigner-Seits radius of $r_s=4$ a.u., the Mie frequency (\ref{Mie}) is 
evaluated as $\omega_0=$3.39 eV. Bertsch and Eckardt fitted the 
electronic density obtained by a self-consistent local density approximation 
by Eq. (\ref{error}) and obtained $a$ =2.14 a.u. for $N=198$ \cite{BE85}. 
Assuming that the surface diffuseness $a$ has a very weak dependense of $N$, 
the energy shift of the double phonon state (\ref{shift}) is estimated as 
\begin{equation}
\Delta^{(2)}E=-0.45 \times N^{-4/3}~~~(eV),
\end{equation}
which is extremely small compared with the phonon frequency $\omega_0$ 
in the harmonic limit. For example, it is as small as $-1.1 \times$ 10$^{-3}$ 
eV for $N$=92. 
This result is consistent with the recent experimental 
observation \cite{SKIH98} that multiple dipole resonances are 
easily accessible for Na clusters. 

In summary, we applied the time-dependent variational principle to 
the dipole plasmon resonance of alkali metal clusters to discuss 
its anharmonic properties. 
The uniform jellium approximation for the ionic density as well as the 
error function approximation for the electronic density 
lead to a simple analytic expression for the energy shift of 
the double phonon state. We found that the ratio of this quantity 
to the frequency in the harmonic limit scales as $N^{-4/3}$, which 
coincides with that for both nuclear giant resonances and 
for the photon spectrum 
in a small cavity. For Na clusters, we found that the anharmonic 
effects are almost negligible, which is consistent with both the 
previous TDLDA calculation and 
the recent experimental suggestion. 

\bigskip

The author is grateful to G.F. Bertsch for useful discussions and 
carefully reading the manuscript. 
This work was supported by the Japan Society for the Promotion of
Science for Young Scientists.

\end{document}